\documentclass[12pt]{article}
\usepackage{epsfig}
\usepackage{amsfonts}
\usepackage{amsopn}
\usepackage{amsmath}

\title{
\vskip -50 pt
\begin{flushright}
\normalsize\rm NORDITA-2011-7
\end{flushright}
\vskip 20 pt Lorentz-invariant membranes \\ 
and finite matrix
approximations}

\author{
Jens Hoppe$^a$ \thanks{e-mail:  hoppe@math.kth.se} \ and  \
Maciej Trzetrzelewski$^{a,b}$ \thanks{e-mail: maciej.trzetrzelewski@gmail.com}  \\
\\
$^a$  Department of Mathematics,\\
Royal Institute of Technology, \\
KTH, 100 44 Stockholm, \\
Sweden \\ \\
$^b$ NORDITA,  \\
Roslagstullsbacken 23, 106 91 Stockholm, \\
Sweden
}

\begin{document}
\date{}
\maketitle

\abstract{The question of Lorentz invariance for finite $N$ approximations of relativistic membranes is addressed. We find that one of the classical manifestations of
Lorentz-invariance is \emph{not} possible for $N \times N$ matrices
(at least when $N=2$ or $3$). How the symmetry is restored in the
large $N$ limit is studied numerically.}

\section{Motivation}
A crucial manifestation of Lorentz invariance in the light-cone
description of the relativistic dynamics of $M$-dimensional extended
objects is the existence of a field $\zeta$ (the
longitudinal embedding coordinate) constructed out of the purely
transverse fields $\vec{x}$ and $\vec{p}$ (and two additional discrete degrees of freedom, $\zeta_0=\int \zeta d^M  \varphi$ and
$\eta=p_{+}$) that satisfies the same  non-linear wave equation then
the components of $\vec{x}$, namely
\begin{equation} \label{1}
\eta^2 \ddot{\zeta}=\Delta \zeta :=\frac{1}{\rho}\partial_a\left(\frac{gg^{ab}}{\rho}\partial_a \zeta\right),
\end{equation}
where $g$ is the determinant of $g_{ab}=\partial_a \vec{x}\partial_b
\vec{x}$,  $\rho$ is a certain non-dynamical density of
unit weight ($\int \rho d^M  \varphi=1$), and the Hamiltonian for the
phase-space variables  $(\vec{x},\vec{p};\eta,\zeta_0)$, constrained
by
\begin{equation} \label{2}
\int f^a \vec{p}\partial_a \vec{x}=0 \ \ \ \hbox{whenever} \ \ \ \partial_a(\rho f^a)=0,
\end{equation}
is given by (cp. \cite{h1,h2})
\begin{equation} \label{3}
H=\frac{1}{2\eta}\int \frac{\vec{p}^2+g}{\rho}d^M \varphi.
\end{equation}
The equations of motion,
\[
\eta \dot{\vec{x}}=\frac{\vec{p}}{\rho}, \ \ \ \eta\frac{\dot{\vec{p}}}{\rho}=\frac{1}{\rho}\partial_a\left(\frac{gg^{ab}}{\rho}\partial_a \vec{x}\right)= \Delta \vec{x}
\]
\begin{equation} \label{4}
\dot{\eta}=0, \ \ \  \dot{\zeta_0}=\frac{H}{\eta},
\end{equation}
and (\ref{2}), imply that $\zeta$ can, via
\begin{equation} \label{5}
\eta^2 \dot{\zeta}=\frac{\vec{p}^2+g}{2\rho^2}, \ \ \ \eta \partial_a \zeta= \frac{\vec{p}}{\rho}\partial_a \vec{x},
\end{equation}
consistently be reconstructed; and (\ref{1}) easily follows from
(\ref{5}):
\[
\eta^2 \ddot{\zeta}= \frac{\vec{p}}{\rho}\frac{\dot{\vec{p}}}{\rho}+\frac{g}{\rho^2}g^{ab}\partial_a \vec{x}\partial_b \dot{\vec{x}}=
\frac{\vec{p}}{\eta\rho}\Delta \vec{x}+ \frac{g}{\rho^2}g^{ab}\partial_a \vec{x}\partial_b \frac{\vec{p}}{\rho \eta}
\]
\begin{equation}  \label{6}
\Delta \zeta = \frac{1}{\rho} \partial_a \left(\frac{g g^{ab}}{\rho}
\partial_b \zeta \right)= \frac{\vec{p}}{\eta \rho} \Delta \vec{x} +
\frac{g}{\rho^2}g g^{ab} \partial_a \vec{x} \partial_b
\frac{\vec{p}}{\eta \rho}.
\end{equation}
Note that the original, manifestly Lorentz-invariant,  formulation,
namely $\bold{\Delta} x^{\mu}=0$ ($x^{\mu}, \ \mu=0,1,\ldots, D-1$ being
the embedding coordinates of the  $M+1$ dimensional manifold
$\mathcal{M}$ swept out in space-time), directly implies (\ref{1}),
as $\zeta=x^0-x^{D-1}$ (and time $\tau =\frac{x^0+x^{D-1}}{2}$), while the
chosen light-cone gauge (cp. \cite{h1,h2}) with $G_{0a}=0$ ($a=1,\ldots,M$),
$G_{ab}=-g_{ab}$ reduces $\bold{\Delta}=\frac{1}{\sqrt{G}}
\partial_{\alpha} \sqrt{G}G^{\alpha\beta}\partial_{\beta}$, the
Laplacian on $\mathcal{M}$, to a non-linear wave operator proportional to $\eta^2
\partial_t^2-\Delta$. Also note that an explicit formula for $\zeta$ was given by
Goldstone in the mid-eighties \cite{h2},
\begin{equation} \label{7}
\zeta=\zeta_0+ \frac{1}{\eta}\int G( \varphi,\tilde{ \varphi}) \tilde{\nabla}^a \left( \frac{\vec{p}}{\rho}\tilde{\partial}_a \vec{x} \right)\rho(\tilde{\varphi}) d^M\tilde{\varphi}
\end{equation}
\[
\int G( \varphi,\tilde{ \varphi})\rho( \varphi)d^M \varphi =0, \ \ \ \Delta_{\tilde{ \varphi}}G( \varphi,\tilde{ \varphi})=\frac{\delta( \varphi,\tilde{ \varphi})}{\rho( \varphi)}-1,
\]
and recently \cite{h3} rewritten as
\[
\zeta_0 + \frac{\vec{x}\cdot\vec{p}}{2\eta\rho}-\int \frac{\vec{x}\cdot \vec{p}}{2\eta \rho}\rho d^M\varphi
\]
\begin{equation} \label{8}
+ \frac{1}{2}\int G( \varphi,\tilde{ \varphi}) \left( \frac{\vec{p}}{\eta \rho}\Delta \vec{x}- \vec{x}\Delta \frac{\vec{p}}{\eta \rho} \right)(\tilde{ \varphi}) \rho(\tilde{ \varphi}) d^M \tilde{ \varphi}
\end{equation}
-implying that $\tilde{\zeta}$, defined as the part of $2\eta(\zeta-\zeta_0)$
not containing $\vec{P}=\int \vec{p}d^M  \varphi$, can be  rewritten as
\begin{equation} \label{9}
\tilde{\zeta}= (d_{\alpha\beta\gamma}+e_{\alpha\beta\gamma})\vec{x}_{\beta}\cdot\vec{p}_{\gamma} Y_{\alpha}( \varphi)
\end{equation}
where
\begin{equation} \label{10}
d_{\alpha\beta\gamma}=\int Y_{\alpha}Y_{\beta}Y_{\gamma}\rho d^M \varphi, \ \ \  e_{\alpha\beta\gamma}:=\frac{\mu_{\beta}-\mu_{\gamma}}{\mu_{\alpha}}d_{\alpha\beta\gamma},
\end{equation}
with $Y_{\alpha}$ and $-\mu_{\alpha}$ being the (non-constant)
eigenfunctions, resp. eigenvalues, of a Laplacian on the parameter
space (with a metric whose determinant is $\rho^2$). Note that
(\ref{7}) satisfies the first equation in (\ref{5}) without having
to use (\ref{2}) (i.e. "strongly")
\[
2\eta^2 \dot{\zeta}=2\eta H + 2\int
G( \varphi,\tilde{ \varphi})\tilde{\nabla}^a \left(
\frac{\vec{p}}{\rho}\partial_a \frac{\vec{p}}{\rho}+\Delta \vec{x}
\partial_a \vec{x} \right)(\tilde{\varphi})\tilde{\rho}d^M  \varphi
\]
\[
= 2\eta H + \int G \tilde{\Delta} \left( \frac{\vec{p}^2}{\rho^2} \right)\rho d^M\varphi+
2\int G( \varphi,\tilde{ \varphi})\tilde{\nabla}^a \left[ \frac{1}{\rho}\partial_b
\left(\frac{gg^{bc}}{\rho}\partial_c \vec{x}\right)
\partial_a\vec{x}  \right]\tilde{\rho} d^M\tilde{\varphi}
\]
\begin{equation} \label{11}
=2\eta H + \int (\tilde{\Delta} G) \left( \frac{\vec{p}^2}{\rho^2} + \frac{g}{\rho^2} \right)(\tilde{\varphi})\rho d^M\tilde{\varphi} = \frac{\vec{p}^2+g}{\rho^2}
\end{equation}
- having used that
\begin{equation} \label{12}
2 \nabla^a \left[\frac{1}{\rho}\partial_b\left(\frac{g}{\rho}\right)\delta^b_a- \frac{1}{2\rho^2}\partial_a g\right]=\Delta \left(\frac{g}{\rho^2}\right).
\end{equation}
On the other hand,
\begin{equation} \label{13}
\eta \partial_a \zeta - \frac{\vec{p}}{\rho}\partial_a\vec{x}=-\int
\left(\partial_a \tilde{\partial}^bG( \varphi,\tilde{ \varphi})+
\frac{\delta( \varphi,\tilde{ \varphi})}{\rho}\delta_a^b\right)\vec{p}\partial_b
\vec{x}(\tilde{\rho})d^M \tilde{ \varphi}
\end{equation}
is of the form (\ref{2}) with
\begin{equation} \label{14}
\tilde{\partial}_b \left[ \tilde{\rho}\left(\partial_a
\tilde{\partial}^bG( \varphi,\tilde{ \varphi})+
\frac{\delta( \varphi,\tilde{ \varphi})}{\rho}\delta_a^b \right) \right] =
\tilde{\rho}\partial_a \tilde{\Delta} G( \varphi,\tilde{ \varphi})- \partial_a
\delta( \varphi,\tilde{ \varphi})=0,
\end{equation}
implying that (\ref{13}) vanishes on the constrained phase space,
hence the second part of (\ref{5}) holds (weakly) too. These
considerations will later when we have to guess/know which part of
$\eta^2 \ddot{\zeta}-\Delta \zeta$ is only weakly zero, of some
relevance.

(\ref{1}) (together with the first equation in (\ref{5})), immediately implies that the Lorentz-generator
\begin{equation} \label{16}
M_{i-}=\int (x_i \mathcal{H}- p_i \zeta)d^M \varphi
\end{equation}
Poisson-commutes with $H$, as
\begin{equation} \label{17}
\eta \dot{M}_{i-}= \int p_i \left(\frac{\mathcal{H}}{\rho}-\eta \dot{\zeta}\right)d^M \varphi   + \int x_i\left(\frac{\dot{\mathcal{H}}\eta}{\rho}-\Delta \zeta\right)\rho d^M \varphi.
\end{equation}

\section{Matrix approximation, $M=2$}

In \cite{h4} the question of Lorentz-invariance of Matrix Membranes  was
discussed and a discrete analogue of $\tilde{\zeta}$ proposed,
\begin{equation} \label{15}
\zeta_N:= \sum_{a,b,c=1}^{N^2-1}\frac{\mu_a+\mu_b-\mu_c}{\mu_a} d_{abc}^{(N)}\vec{x}_b\cdot\vec{p}_c T_a^{(N)}
\end{equation}
where the $T_a^{(N)}$ are hermitean $N \times N$ matrices and
$d_{abc}^{(N)}$ is proportional to $Tr(T_a^{(N)}\{T_b^{(N)},T_c^{(N)}\})$ (the
normalizations suited for $N \to \infty$ will be discussed below).

In this paper we would like to address the question of
Lorentz-invariance for the Matrix theory, focusing on numerical
computations, that will tell us that
\begin{itemize}
\item at least for low $N$ (probably
all finite $N$) there are \emph{no} Matrix solutions for the natural
analogue of (\ref{1})
\item show how exactly (and how not) the finite $N$ analogue (\ref{15}) will approach "solving (\ref{1})".
\end{itemize}

Before going into the details, let us note that, on general grounds (cp. \cite{h45,h5}) one is guaranteed that
for $M=2$ and any genus, a sequence $T_{\alpha}$,
$\alpha=1,2,\ldots$ of linear maps (from real functions to hermitean
matrices)
\begin{equation} \label{18}
T_{\alpha}: f \to F^{(N_{\alpha})}=T_{\alpha}(f)
\end{equation}
exists, as well as a sequence of increasing positive integers
$N_{\alpha}$, and decreasing positive real numbers $\hbar_{\alpha}$
(with $ \lim_{\alpha \to \infty}\hbar_{\alpha}N_{\alpha}$ finite),
with the following properties (for arbitrary smooth functions
$f,g,\ldots$):
\[
\lim_{\alpha \to \infty} || T_{\alpha}(f) || < \infty,
\]
\[
|| T_{\alpha}(f)T_{\alpha}(g) - T_{\alpha}(f \cdot g) || \to  0,
\]
\[
|| \frac{1}{i\hbar_{\alpha}}[T_{\alpha}(f),T_{\alpha}(g)] - T_{\alpha}(\{f,g\}) || \to  0,
\]
\begin{equation} \label{19}
2\pi \hbar_{\alpha} Tr(T_{\alpha}(f)) \to  \int f \rho d^2  \varphi.
\end{equation}
Here $||T_{\alpha}(f)||$ can be taken as  the largest eigenvalue of
the hermitean matrix $T_{\alpha}(f)$, and
\begin{equation} \label{20}
\{f,h\}:= \frac{\epsilon^{rs}}{\rho}\partial_r f \partial_s h.
\end{equation}
For functions on a sphere this map exists for all integers $N>1$
(hence one can drop the index $\alpha$ and simply write $N$ and
$\hbar_N$ instead of $N_{\alpha}$ and $\hbar_{\alpha}$) and, up to
normalisation is given \cite{h1} via replacing in
\begin{equation} \label{21}
Y_{lm}(\theta, \varphi) = \sum_{A_k=1,2,3} c^{(lm)}_{A_1 \ldots A_l}x_{A_1}\ldots x_{A_l}|_{\vec{x}^2=1},
\end{equation}
the commuting variables
\[
x_1=r \sin\theta\cos \varphi, \ \ x_2=r\sin\theta\sin \varphi, \ \ x_3=r\cos\theta
\]
by three $N\times N$ matrices $X_1$, $X_2$, $X_3$ satisfying
\[
[X^{(N)}_A,X^{(N)}_B]=i \frac{2}{\sqrt{N^2-1}} \epsilon_{ABC}X^{(N)}_C,
\]
\begin{equation} \label{22}
\vec{X}^2:=X_1^2+X_2^2+X_3^2=\bold{1}_{N\times N}.
\end{equation}
The resulting "Matrix harmonics" $T_{lm}=T_N(Y_{lm})$ (linear
independent for $l=0,1,\ldots,N-1$, $m=-l,\ldots,+l$, and
identically zero for $l \ge N$)  are known \cite{h1,h2,h6} to have many
special properties. In particular, they are eigenfunctions of the
discrete Laplacian with eigenvalues $-\mu_{lm}$ being equal to the
infinite $N$ eigenvalues $-l(l+1)$ of the parameter space Laplace
\begin{equation} \label{23}
\Delta = \frac{1}{\sin \theta} \partial_{\theta} \sin\theta \partial_{\theta} + \frac{1}{\sin^2\theta}\partial_ \varphi^2.
\end{equation}

Apart from the fact that those $T_{lm}^{(N)}$ are not hermitean (because of the $Y_{lm}$ being complex), they are ideally suited for
testing (\ref{15}). Note that for an arbitrary function $f$ the map
$T_N$ can be explicitly given as
\begin{equation} \label{24}
f=\sum_{l=0, |m|\le l}^{\infty} f_{lm}Y_{lm}(\theta,  \varphi)  \to F^{(N)} := \sum_{l=0}^{N-1} f_{lm}T_{lm}^{(N)}.
\end{equation}

To get the normalisation factors right is more difficult for a variety of reasons:
from a practical/computational point of view it is easiest to take
$\hat{T}_{lm}^{(N)}$ as given in \cite{h6}, satisfying
$Tr(\hat{T}_{lm}^{(N)}\hat{T}_{l'm'}^{(N)})=\delta_{ll'}\delta_{mm'}$, and in
particular equation (6) in \cite{h6} as well as
\begin{equation} \label{25}
\hat{T}_{lm}^{(N)} = \sqrt{4\pi}\sqrt{\frac{(N^2-1)^l(N-1-l)!}{(N+l)!}}\sum_{A_k=1,2,3} c^{(lm)}_{A_1\ldots A_l}X^{(N)}_{A_1}\ldots X^{(N)}_{A_l}.
\end{equation}
Apart from further "hermiteanization" note that the usual spherical harmonics are normalized according
to $\int Y_{lm}^*Y_{l'm'}\sin\theta d\theta
d \varphi=\delta_{ll'}\delta_{mm'}$ whereas $\rho$ should satisfy $\int
\rho d^2  \varphi=1$. Hence with $\rho \to \frac{\rho}{4\pi}$, $Y_{lm}\to
\sqrt{4\pi}Y_{lm}$, explaining the factor $\sqrt{4\pi}$ in
(\ref{25}).

To conform with (\ref{19}) we need $N 2\pi \hbar_N \to 1$; we
choose\footnote{One could also take $2\pi N \hbar=1$ (for
all $N$) and(or) multiply $\hat{T}$ by $\sqrt{N}$, rather then
(cp. (\ref{28})) by $(N^2-1)^{\frac{1}{4}}$. This would have the
advantage of having $T(1)=\bold{1}$ and $T(x_A)=X_A$ hold exactly,
for any finite $N$, and also simplify (\ref{29}). 
}
\begin{equation} \label{26}
\hbar_N=\frac{1}{2\pi\sqrt{N^2-1}}.
\end{equation}
In order to have
\begin{equation} \label{27}
2\pi \hbar_N Tr(T^{\dagger \ (N)}_{lm}T_{l'm'}^{(N)}) \to \delta_{ll'}\delta_{mm'}
\end{equation}
we multiply (see previous footnote) (\ref{25}) by $(N^2-1)^{\frac{1}{4}}$ i.e. take
\begin{equation} \label{28}
\tilde{T}_{lm}^{(N)}= (N^2-1)^{\frac{1}{4}} \hat{T}_{lm}^{(N)}= \sqrt{4\pi}\gamma_{Nl}\sum_{A_k=1,2,3} c^{(lm)}_{A_1 \ldots A_l}X_{A_1}\ldots X_{A_l}
\end{equation}
with $\gamma_{Nl}\to 1$ as $N \to \infty$.

Due to $\mu_{lm}=\mu_{l-m}$, and $T^{\dagger}_{lm}\sim
(-1)^mT_{l-m}$ (as a consequence of $Y^*_{lm}=(-1)^m Y_{lm}$) one
can at the end easily get rid of the non-hermicity problem and
consider hermitean matrices $T_a^{(N)}$. Forming linear combinations
$\frac{T_{lm}^{(N)}+T^{\dagger \ (N)}_{lm}}{\sqrt{2}}$ and
$\frac{T_{lm}^{(N)}-T^{\dagger \ (N)}_{lm}}{\sqrt{2}i}$ one obtains the desired hermitean
basis $\{T_a^{(N)}\}_{a=0}^{N^2-1}$ satisfying
\begin{equation} \label{29}
\frac{1}{\sqrt{N^2-1}}Tr\left(T^{(N)}_aT^{(N)}_b\right)=\delta_{ab}.
\end{equation}
The matrix approximation of (\ref{3}) is then given by  (leaving out
$\eta$ from now on, which - just as is done in string theory - can,
for most purposes, be absorbed in a redefinition of "time")
\[
H_N= \frac{1}{2\sqrt{N^2-1}}Tr\left(\vec{P}^2-\frac{1}{2}(2\pi)^2(N^2-1)[X_i,X_j]^2\right)
\]
\begin{equation} \label{30}
=\frac{1}{2}\left(p_{ia}p_{ia}+\frac{1}{2}f^{(N)}_{abc}f^{(N)}_{ab'c'}\vec{x}_b \cdot \vec{x}_{b'} \vec{x}_c \cdot \vec{x}_{c'}\right)
\end{equation}
and the normalisations, and conventions,
\begin{equation} \label{31}
d_{abc}^{(N)}:= \frac{1}{2\sqrt{N^2-1}}Tr\left(T^{(N)}_a\{T^{(N)}_bT^{(N)}_c\}\right)
\end{equation}
(the symbol $\{\cdot,\cdot\}$ here denotes the anticommutator of matrices) and
\begin{equation} \label{32}
f_{abc}^{(N)}:= \frac{2\pi}{i}Tr\left(T^{(N)}_a[T^{(N)}_b,T^{(N)}_c]\right),
\end{equation}
are such that, as $N\to \infty$, (\ref{31}) and (\ref{32}) approach, respectively,
\begin{equation} \label{33}
d_{abc}= \int Y_aY_bY_c \rho d\theta d \varphi, \ \ \ g_{abc}= \int Y_a\{Y_b,Y_c\}\rho d\theta d \varphi, \ \ \ \rho=\frac{\sin\theta}{4\pi}.
\end{equation}

\section{Lorentz symmetry at finite $N$?}
We now focus on calculating $\ddot{\zeta}_N-\Delta^{(N)} \zeta_N$ where we take $\zeta_N$  to be given by
\[
\zeta_N= \sum_{a,b,c=1}^{N^2-1}L_{abc}^{(N)}\vec{x}_b \vec{p}_c T_a^{(N)}
\]
with for the moment \emph{arbitrary} coefficients $L_{abc}^{(N)}$. Using the discrete equations of motion $\dot{x}_{ia}=p_{ia}$, $\dot{p}_{ia}=f_{abc}^{(N)}f_{c'ba'}^{(N)}\vec{x}_c\cdot \vec{x}_{c'} x_{ia'}=\Delta^{(N)}_{aa'}x_{ia'}$ 
we find that
\begin{equation} \label{34}
\ddot{\zeta}_N-\Delta^{(N)} \zeta_N = \vec{x}_m\cdot\vec{x}_n \ \vec{x}_{\mu}\cdot\vec{p}_{\nu}R_{a m n \mu \nu}^{(N)}T_a^{(N)}
\end{equation}
where
\[
R_{a m n \mu \nu}^{(N)}=L_{ac\nu}^{(N)}f_{cdm}^{(N)}f_{nd\mu}^{(N)} +
L_{a\mu c}^{(N)}f_{cdm}^{(N)}f_{nd\nu}^{(N)}+
2L_{a\nu c}^{(N)}f_{cdm}^{(N)}f_{nd\mu}^{(N)}
\]
\begin{equation} \label{35}
+L_{amc}^{(N)}(f_{cd\mu}^{(N)}f_{\nu dn}^{(N)}+f_{cd\nu}^{(N)}f_{\mu dn}^{(N)}) -
L_{c\mu\nu}^{(N)}f_{adm}^{(N)}f_{ndc}^{(N)}.
\end{equation}
The question that now arises is whether there exist nontrivial
coefficients $L_{abc}^{(N)}$ (i.e. which cannot be written as
$M_{ak}^{(N)}f_{kbc}^{(N)}$) such
that the r.h.s of equation (\ref{34}) is weakly zero. The
corresponding equation for $R_{a m n \mu \nu}^{(N)}$ is
\begin{equation} \label{36}
R_{a m n \mu \nu}^{(N)}+R_{a n m \mu \nu}^{(N)} = G_{anmk}^{(N)}f_{k\mu\nu}^{(N)}
\end{equation}
where $G_{anmk}^{(N)}$ are unknown coefficients. Note that $M_{ak}^{(N)}f_{kbc}^{(N)}$ is a solution of (\ref{36}) for arbitrary $M_{ab}$, i.e.
satisfies (\ref{36}) with
\[
G_{anmk}^{(N)} = - M_{ck}^{(N)}(f_{amd}^{(N)}f_{dnc}^{(N)}+f_{and}^{(N)}f_{dmc}^{(N)}).
\]

In order to see that there are no other solutions it is best to
first symmetrize (\ref{36}) over $\mu$ and $\nu$
\[
R_{a m n \mu \nu}^{(N)}+R_{a n m \mu \nu}^{(N)} +R_{a m n \nu \mu}^{(N)}+R_{a n m \nu \mu}^{(N)}=0
\]
and solve the resulting equation with respect to $L_{abc}^{(N)}$.
By explicit calculation for $N=2$ and $N=3$ we found the general solution to be of the form
$L_{abc}^{(N)}=M_{ak}^{(N)}f_{kbc}^{(N)}$, proving that nontrivial solutions do not exist (for $N=2,3$, possibly for all $N$).

On the other hand, in the large $N$ limit the
theory is relativistically invariant \cite{h2}; therefore there should exist a
choice of $L_{abc}^{(N)}$ such that the r.h.s. of (\ref{34})
converges to zero when $N \to \infty$ for $\vec{x}_a$ and
$\vec{p}_a$  satisfying the Gauss constraint
$G_a:=f_{abc}^{(N)}\vec{x}_b \vec{p}_c=0$. In the following we will
consider $\zeta_N$ given by (\ref{15}) i.e. we take
\begin{equation} \label{388}
L_{abc}^{(N)}=\frac{\mu_a+\mu_b-\mu_c}{\mu_a} d_{abc}^{(N)}.
\end{equation}
Inserting (\ref{388}) into (\ref{34}) however does not immediately
yield a r.h.s. converging to zero (as we found numerically); hence it
is necessary to explicitly determine $G^{(N)}_{anmk}$. To derive the
exact form of the subtraction that one has to make to render convergence
(to zero), as $N \to \infty$, is non-trivial:
due to
\[
\Delta \zeta: = \frac{1}{\rho}\partial_b \left(\frac{gg^{ab}}{\rho}\partial_a \zeta \right),
\]
the term involving constraints is (leaving out the $\rho$-factors and $\eta$ for simplicity), cp. (\ref{13}):
\[
-\partial_b\left(gg^{ab}\int F^c_a(\varphi,\tilde{\varphi})(\vec{p}\partial_c \vec{x})(\tilde{\varphi})d^M\varphi  \right);
\]
for $M=2$, the $\alpha$-component of that is
\[
+\int\int\left( \partial_b Y_{\alpha}\epsilon^{bb'}\epsilon^{aa'}\partial_{a'}\vec{x}\partial_{b'}\vec{x} \right)(\varphi) F^c_a(\varphi,\tilde{\varphi})(\vec{p}\partial_c \vec{x})(\tilde{\varphi})d^2 \varphi d^2 \tilde{\varphi}
\]
\[
=-\sum_{\gamma}\int \{Y_{\alpha},\vec{x}\}\partial_a \vec{x}\frac{\partial^a Y_{\gamma}}{\mu_{\gamma}}\{\vec{x},\vec{p}\}_{\gamma}=-\frac{1}{2}g_{\alpha m \beta}g_{\gamma\mu\nu}L_{\gamma n\beta}\vec{x}_m \cdot \vec{x}_n \vec{x}_{\mu} \cdot \vec{p}_{\nu}
\]
as, using the completeness of vector spherical harmonics on $S^2$
\[
F^c_a:=\sum_{\gamma=1}^{\infty}-\frac{1}{\mu_{\gamma}}\left( \partial_a Y_{\gamma}(\varphi) \tilde{\partial}^c Y_{\gamma}(\tilde{\varphi})\right)+\delta^a_c \delta(\varphi,\tilde{\varphi})
= \sum_{\gamma=1}^{\infty} \frac{\epsilon_{aa''}}{\mu_{\gamma}}\partial^{a''}Y_{\gamma}(\varphi)\epsilon^{cc'}\partial_{c'}Y_{\gamma}(\tilde{\varphi}).
\]
Hence (note the factor of $2$ involved in the relation between $\zeta$ and $\tilde{\zeta}$) the matrix
\begin{equation} \label{39}
U:=\vec{x}_m\cdot\vec{x}_n \ \vec{x}_{\mu}\cdot\vec{p}_{\nu}(R_{a m n \mu \nu}^{(N)}-S_{amn\mu\nu}^{(N)})T_a^{(N)}
\end{equation}
with
\[
S_{amn\mu\nu}^{(N)}:=-L_{cnd}^{(N)}f_{adm}^{(N)}f_{c\mu\nu}^{(N)}
\]
should not contain any terms proportional to the $G_a$'s and therefore should converge strongly to $0$ in the large $N$ limit.

\subsection*{Numerical investigation}

In order to verify that the matrix $U$ indeed converges to $0$ we
performed a numerical analysis for matrices with $N=3,\ldots,11$
using the conventions described in section 2 ($N=2$ is trivial, $U=0$). The elements of the
matrix $U$ are polynomials of the form
 \[
U^{(N)}_{ij}:=\vec{x}_m\cdot\vec{x}_n \ \vec{x}_{\mu}\cdot\vec{p}_{\nu}\tilde{R}_{a m n \mu \nu}^{(N)}[T_a^{(N)}]_{ij}
 \]
where $\tilde{R}_{a m n \mu \nu}^{(N)}:=R_{a m n \mu \nu}^{(N)}-S_{amn\mu\nu}^{(N)}$. We restrict the analysis to $i,j=1,2,3$, i.e.
we analyze what is the $N$ dependence of the $SU(3)$ corner of
matrix $U$, and to $1 \le a,m,n,\mu,\nu \le 8$, i.e. we consider only
the range of the $SU(3)$ adjoint index.

A typical polynomial $U^{(N)}_{ij}$ consists of about 700 terms satisfying
these restrictions. We found numerically that they all behave like
$1/N$ (see Fig. 1).

\begin{figure}[h]
\centering
\leavevmode
\epsfverbosetrue
\epsffile{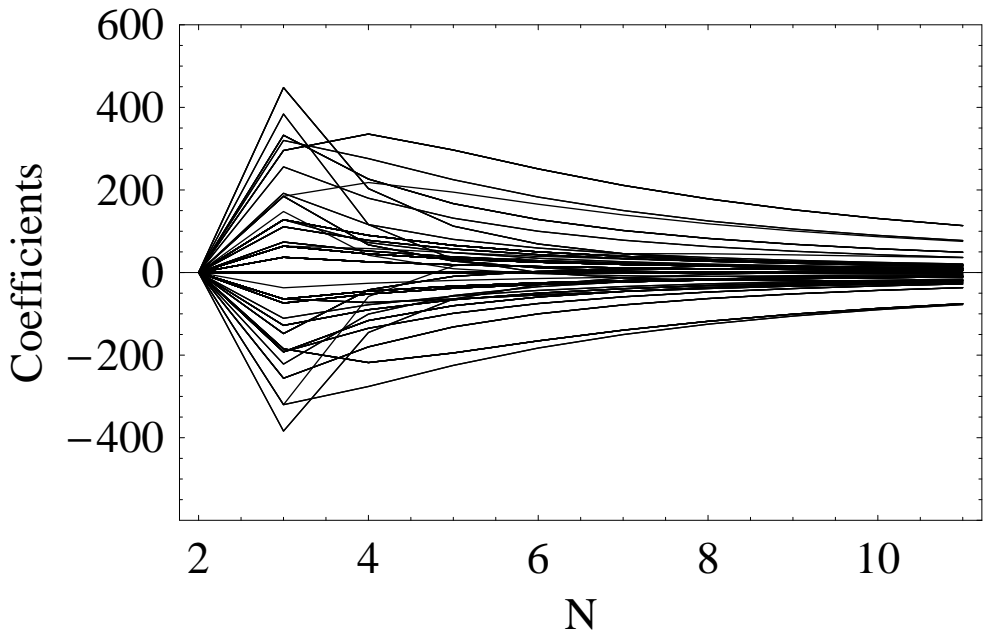}
\caption{$N$ dependence of coefficients from $U^{(N)}_{ij}$ polynomials  }
\end{figure}

We would like to make several comments concerning this result.
First, the fact that we subtracted the Gauss constraint by
considering $\tilde{R}$ instead of $R$ is necessary to see the
convergence. If the Gauss constraint is not subtracted then the
corresponding polynomials $U^{(N)}_{ij}$ diverge - their
coefficients behave like $N^1$. Second, the combinations of terms in
(\ref{35}) is of course very special i.e. crucial for the convergence. If for
instance we consider only the first term in (\ref{35}), i.e.
$L_{ac\nu}^{(N)}f_{cdm}^{(N)}f_{nd\mu}^{(N)}$ then the coefficients
of the resulting polynomial  $U^{(N)}_{ij}$ are divergent,
behaving like $N^1$. Third, the restrictions ($1 \le i,j \le 3$ and $1
\le a,m,n,\mu,\nu \le 8$)  we used are certainly minimal. The question
remains to what extent one can relax these restrictions still having
the convergence. It is reasonable to conjecture that for any fixed
$n<N$ (i.e. $n$ independent of $N$) the elements of the matrix
$U^{(N)}_{ij}$ satisfying  the restrictions $1 \le i,j \le n, \ \ \  1 \le a,m,n,\mu,\nu \le n^2-1$
still converge to $0$. \\

\vspace{12pt} \noindent{\bf Acknowledgment} Support from the Swedish
Research Council is gratefully acknowledged.


\begin{thebibliography}{7}


\bibitem{h1}
J. Hoppe, \textit{Quantum theory of a massless relativistic surface and a
two-dimensional bound state problem}, PhD Thesis MIT 1982
(http://dspace.mit.edu/handle/1721.1/15717). See also J. Hoppe, \textit{Membranes and
matrix models}, arXiv:hep-th/0206192 (IHES/P/02/47) and references therein.

\bibitem{h2}
J. Goldstone, unpublished.

\bibitem{h3}
J. Hoppe, \textit{Fundamental structures of M(brane) theory}, Phys. Lett. B 695 (2011) 384,
doi:10.1016/j.physletb.2010.11.038
{\tt arXiv:1003.5189 hep-th}.

\bibitem{h4}
J. Hoppe, \textit{Matrix Models and Lorentz Invariance}, J. Phys. A 44 (2011) 055402 doi:10.1088/1751-8113/44/5/055402 {\tt
arXiv:1007.5505 hep-th}.

\bibitem{h45}
M. Bordemann, E. Meinrenken, M. Schlichenmaier, \textit{Toeplitz quantization of K\"ahler manifolds and $gl(N)$, $N \to \infty$ limits}, Comm. Math. Phys. 165 (1994) 281.

\bibitem{h5}
J. Arnlind, J. Hoppe, G. Huisken, \textit{Discrete Curvature and the Gauss-Bonnet Theorem} {\tt arXiv:1001.2223 hep-th}.

\bibitem{h6}
J. Hoppe, S. T. Yau, \textit{Some Properties of Matrix Harmonics on $S^2$} Comm. Math. Phys. 195 (1998) 67.


\end{thebibliography}
\end{document}